\documentclass[conference]{IEEEtran}
\IEEEoverridecommandlockouts
\usepackage{cite}
\usepackage{amsmath,amssymb,amsfonts}
\usepackage{algorithmic}
\usepackage{graphicx}
\usepackage{textcomp}
\usepackage{xcolor}
\usepackage{subcaption}
\usepackage{hyperref} 
\usepackage{booktabs}

\def\BibTeX{{\rm B\kern-.05em{\sc i\kern-.025em b}\kern-.08em
    T\kern-.1667em\lower.7ex\hbox{E}\kern-.125emX}}
\begin{document}

\title{Including Follower Dynamics in Beaconing for Platooning Safety}

\author{\IEEEauthorblockN{ Hassan Laghbi \href{https://orcid.org/0009-0004-3751-9081}{\includegraphics[scale=0.15]{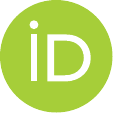}}}

\IEEEauthorblockA{\textit{School of Computing} \\
\textit{Newcastle University}\\
Newcastle upon Tyne, UK \\
h.laghbi2@newcastle.ac.uk \\
\textit{College of Engineering and CS} \\
\textit{Jazan University}\\
Jazan, KSA \\
hlaghbi@jazanu.edu.sa
}
\and
\IEEEauthorblockN{ Nigel Thomas }
\IEEEauthorblockA{\textit{School of Computing} \\
\textit{Newcastle University}\\
Newcastle upon Tyne, UK \\
nigel.thomas@newcastle.ac.uk}
}

\maketitle

\begin{abstract}
In this paper, we propose procedures to address platoon follower dynamics within adaptive beaconing. We implement them in a known adaptive beaconing scheme which is Jerk Beaconing (JB) to improve its safety. We evaluate our proposed approach in terms of safety, string stability and the channel busy ratio (CBR) overhead. The results reveal that our proposal significantly enhances safety without imposing substantial CBR overhead and maintains the string stability of the PATH CACC controller under normal conditions.
\end{abstract}

\begin{IEEEkeywords}
beaconing, platooning, CACC, safety
\end{IEEEkeywords}

\section{Introduction}
Vehicle platooning is an application of Vehicular Ad-hoc Networks (VANETs) which aims to enhance fuel efficiency, particularly for heavy-duty vehicles. Additionally, it increases road capacity by minimising inter-vehicle distances and improves traffic safety by mitigating traffic shock waves which can contribute to accidents \cite{b0}.

Platooning relies on Cooperative Adaptive Cruise Control (CACC) controllers installed in each platoon member. These controllers require input from other vehicles within the platoon which is communicated via a vehicle-to-vehicle (V2V) wireless network using one-hop beacon messages. Beaconing can lead to network congestion which results in packet delays and losses. This, in turn, may prevent the CACC controllers from receiving sufficient input, potentially compromising safety. To address this issue, various adaptive beaconing schemes have been proposed. One recognised approach is Jerk Beaconing (JB), which aims to manage network congestion while maintaining network reliability. However, JB and many adaptive beaconing schemes (e.g., \cite{b6}, \cite{b7}, \cite{b10}, \cite{b12}, \cite{b13}) assume that vehicle dynamics are initiated solely by platoon leaders without accounting for those triggered by platoon followers. Therefore, this paper addresses this limitation by proposing efficient procedures to incorporate follower-initiated dynamics into JB to enhance its safety. The proposed procedures can be integrated into a platooning management or coordination protocol, such as \cite{b16} which includes management and manoeuvring functions such as discovery, formation, joining, leaving, and merging.

\section{Related Work}

Martijn et al. \cite{b5} highlight that CACC controllers ideally should receive between 10 and 25 updates per second. However, this leads to congestion on the shared communication channel when vehicle density is high, especially considering that other applications besides platooning may be running on both platooning and non-platooning vehicles within the same communication range. Therefore, numerous schemes have been proposed to reduce contention or adaptively adjust the beacon generation rate to mitigate congestion.

The ETSI DCC (Distributed Congestion Control) is a standard algorithm based on a state machine with three states (relaxed, active, and restrictive) to adapt the beaconing interval. The transitioning between states is based on measurement of the Channel Busy Ratio (CBR). That is, when the CBR is high, DCC reduces the beaconing rate and when it is low, the rate can be increased. In \cite{b6}, DCC has been shown to excessively restrict beacon frequency leading to negative impacts on performance and safety of platooning. Another algorithm is LIMERIC \cite{b7, b8} which controls congestion by dynamically adjusting each vehicle's beaconing rate to achieve a target collective CBR.

Other algorithms, such as Dynamic Beaconing (DynB) \cite{b9} and Adaptive Beacon Generation Rate (ABGR) \cite{b10}, adaptively adjust beaconing generation based on vehicle density. DynB adapts the beaconing interval based on both the number of one-hop neighbours and current CBR above a certain threshold. ABGR controls the beaconing rate in response to vehicle density which is determined based on the received beacons assessed against a set threshold. When density is low, the beaconing rate stays at its peak, but as more vehicles enter the communication range, the rate decreases to mitigate congestion. All the aforementioned schemes lack awareness of platooning dynamics and adjust beaconing solely based on network parameters without considering the specific requirements of platooning, which can compromise safety.

Other schemes are specifically designed to be aware of platooning requirements while utilising the communication channel efficiently. In \cite{b11}, Segata et al. designed Slotted Beaconing which reduces contention between platoon members using a TDMA (Time Division Multiple Access) overlay for 802.11p. In other words, the beaconing interval is divided into fixed time slots to ensure that each platoon member has a designated slot for transmission which eliminates contention. In \cite{b111}, Slotted Beaconing was extended to include more than a single platoon using grouping. Another TDMA scheme was proposed in \cite{b12} which in addition to that allocates some of the slots to other type of messages (event-driven messages) and incorporates a relay selection functionality as well. Based on Slotted Beaconing, Segata et. al. \cite{b13} proposed a dynamic approach to beaconing called Jerk Beaconing which adapts the beaconing interval based on jerk. To enhance the reliability of the scheme, acknowledgments and retransmissions were incorporated in the scheme as well.

\subsection{Jerk Beaconing}
JB (Jerk Beaconing) \cite{b13} dynamically adjusts the beaconing interval $\Delta_{\text{msg}}$ based on jerk $\Delta u$ which represents the change in acceleration between the current and previous values using formulas \eqref{eq:delta_msg} and \eqref{eq:a}. In a platoon based on JB, vehicle 0 sends a scheduled beacon which prompts vehicle 1’s CACC to compute a new control action which it then broadcasts with updated information (and a piggybacked acknowledgment map) and this slotted process propagates sequentially through the platoon. When the beaconing is happening at the maximum beaconing interval, vehicles can estimate the speed of the lead and front vehicles using previously received data.

\begin{equation}
\label{eq:delta_msg} 
\Delta_{\text{msg}}(\Delta u) = \max \left( e^{-a |\Delta u|^p} \cdot \text{max}_{bi}, \text{min}_{bi} \right)
\end{equation}

\noindent The parameter $p$ determines the responsiveness of the scheme, while the parameter $a$ is defined as follows:

\begin{equation}
\label{eq:a} 
a = -\ln \left( \frac{\min_{b_i}}{\max_{b_i}} \right) \cdot \Delta u_{\max}^{-p}
\end{equation}

\section{Proposed Approach}
JB responds only to acceleration changes initiated by the platoon leader which can pose significant safety risks. To address this, we propose using three types of special beacons (messages) to enhance safety, referring to this improved version as JBE (JB Enhanced). The proposed beacons are not predictive beacons to be sent before an event occurs; rather, they are sent during dynamic changes. Unlike normal beacons that provide only partial status updates, such as current speed or acceleration, the proposed beacons convey the complete intended outcome including the final speed or acceleration.

\begin{figure}[]
    \centering
    \includegraphics[scale=0.5]{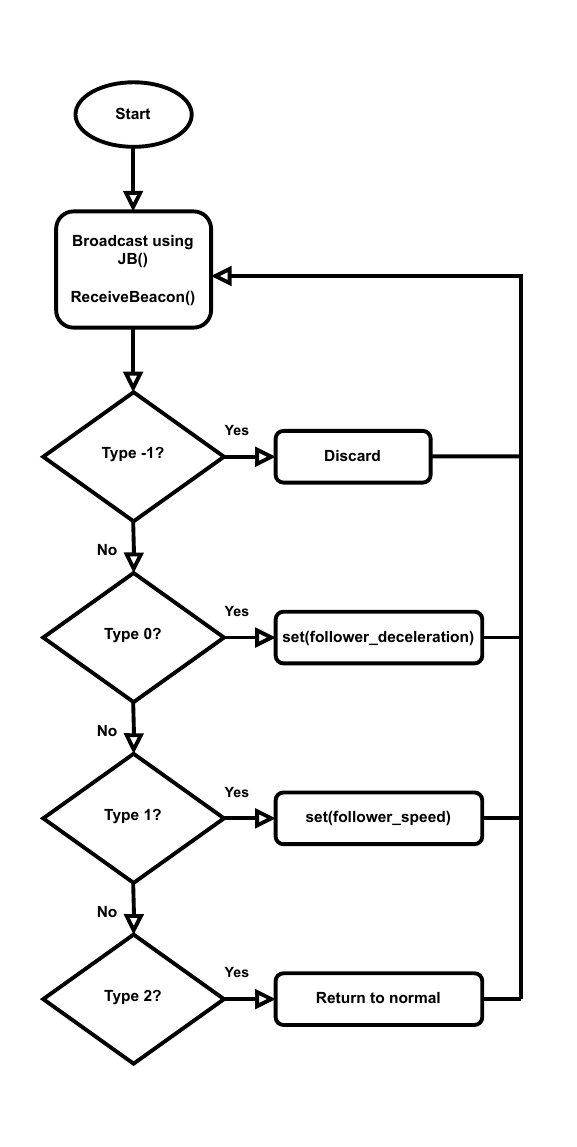}
    \caption{JBE leader loop.}
    \label{fig:jbe_leader_loop}
\end{figure}

Figure \ref{fig:jbe_leader_loop} shows the main JBE loop of a platoon leader. A platoon leader will use JB for beaconing and therefore it will discard any normal beacons (Type -1) from its followers. When it receives a Type 0 beacon, it will extract the deceleration value sent by its follower and apply it immediately, and continue beaconing using JB. This type indicates an emergency braking to a complete stop. If it receives a beacon of Type 1, it will extract the desired speed value sent by its follower and apply it immediately, and continue beaconing using JB. Type 1 indicates slowing down to a new target speed. Upon receiving a Type 2 beacon, the leader reverts to its state before considering follower dynamics, restoring its original desired speed.

\begin{figure}[]
    \centering
    \includegraphics[scale=0.5]{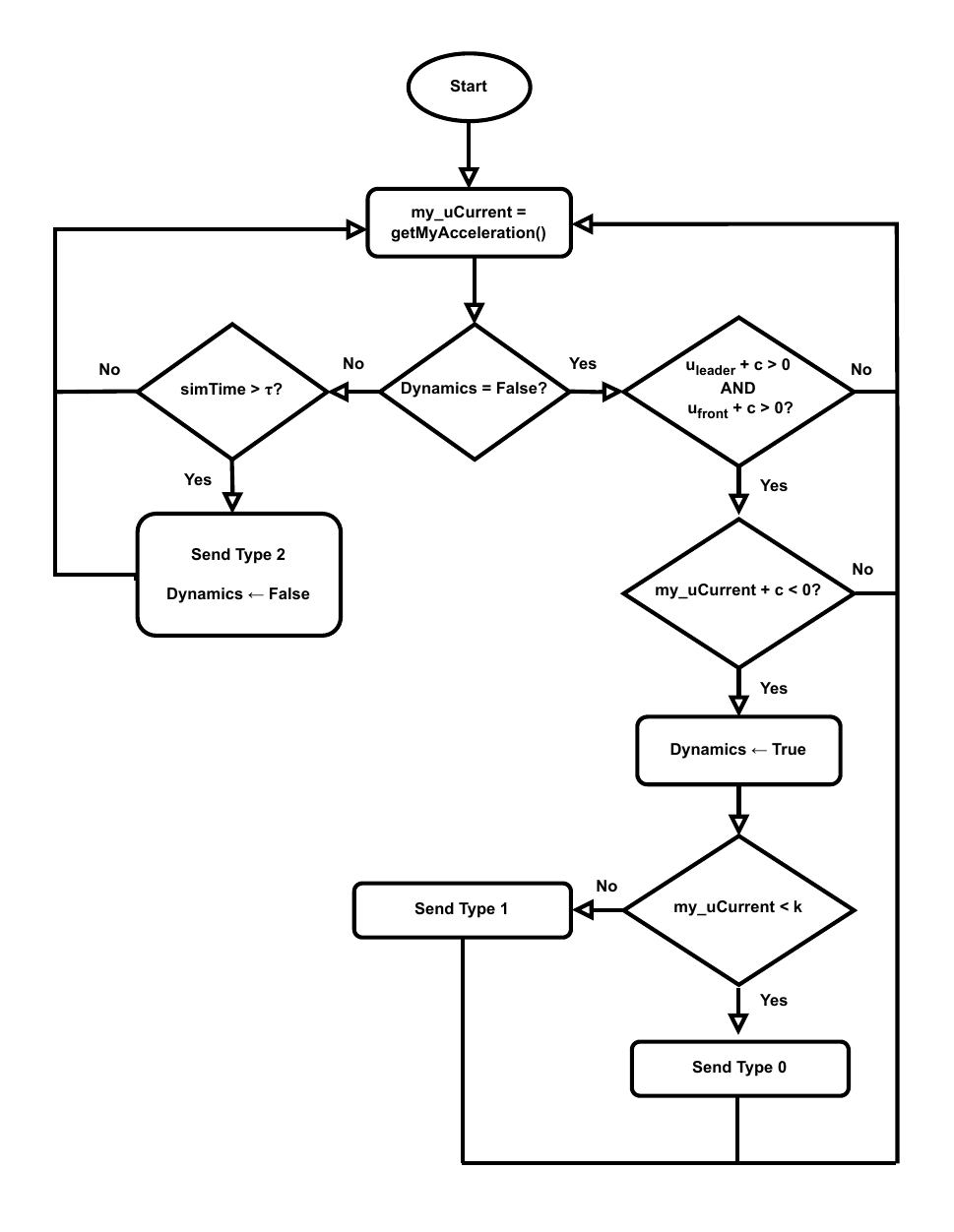}
    \caption{JBE follower loop.}
    \label{fig:jbe_follower_loop}
\end{figure}

A platoon follower also broadcasts beacons using JB and discards any received beacons that are not of Type -1 to avoid responding to the dynamics of other followers. Furthermore, each follower monitors its own dynamics and sends relevant beacon types accordingly. Figure \ref{fig:jbe_follower_loop} shows the main JBE loop of a platoon follower. Each follower continuously monitors its current acceleration. If it is not already in a dynamics state, it checks the most recent acceleration values received from both the platoon leader and the vehicle directly ahead. This enables the follower to determine whether its dynamics are merely a reaction to the leader or the preceding vehicle, or if they were initiated by the follower itself. In the flowchart, the constant $c$ serves as an offset to discard slight decelerations. If neither the leader nor the vehicle in front is decelerating, the follower checks whether it is decelerating. Acceleration is disregarded, as we assume followers are not malicious and do not accelerate beyond the leader's acceleration. It is assumed that a follower would only need to decelerate or come to a complete stop. If a follower is decelerating, it compares its deceleration to a threshold value $k$ (e.g., -5 $\text{m/s}^2$ ) to determine whether to send a Type 0 or Type 1 beacon. If the deceleration exceeds the threshold, it sends a Type 0 beacon, instructing the leader to stop. Otherwise, it sends a Type 1 beacon, including the required speed to be set by the leader. Once in a dynamics state, at any time, a follower can send a Type 2 beacon to the leader to instruct it to revert to its original state returning to its initial speed prior to the slowdown or stop.

\section{System model}
The system under consideration is similar to the one we used in our previous work \cite{b17} which consists of a four-lane highway with a number of platoons, each consisting of 15 vehicles. The first vehicle in each platoon acts as the leader while the remaining 14 are followers. Each follower uses PATH CACC as its platooning controller. The controller receives input data—speed, position, and acceleration—from both the platoon leader and the preceding vehicle to calculate the necessary acceleration or deceleration, which is then fed to a lower-level controller for throttle or braking. The control law for PATH CACC is as follows \cite{b3}:

\begin{equation}
\label{eq:equation1} 
\begin{split} 
u_i = &\ \alpha_1 u_{i-1} + \alpha_2 u_0 + \alpha_3 (-d_{radar} + d_d) \\ &\ + \alpha_4 (\dot{x}_i - \dot{x}_0) + \alpha_5 (\dot{x}_i - \dot{x}_{i-1}), 
\end{split} 
\end{equation}

\noindent where \(i\) represents the index of a vehicle, and \(\dot{x}_i\) denotes its speed. \({u}_{i-1}\) and \(\dot{x}_{i-1}\) denotes the acceleration and speed of the vehicle ahead. Likewise, \(u_0\) and \(\dot{x}_0\) denotes the acceleration and speed of the platoon leader. All acceleration and speed values are acquired through a V2V wireless communication network as all the vehicles are equipped with IEEE802.11p-based on-board units. The radar-detected distance to the preceding vehicle is given by \(d_{radar}\). The constants \(\alpha_i\) are tunable parameters. 

Each platoon member has a platooning application operating at the application layer which sends beacons based on the selected beaconing scheme (JB or JBE). When a new beacon is received from either the platoon leader or the preceding vehicle, the application extracts the necessary data—such as current position, speed, and acceleration—and forwards it to the CACC controller. Table \ref{table1} shows a complete list of the used simulation parameters.

\begin{table}[t]
\centering
\caption{Simulation Parameters}
\label{table1}
\resizebox{0.47\textwidth}{!}{ 
\begin{tabular}{@{}ll@{}}
\toprule
\textbf{Parameter}                                      & \textbf{Value}                    \\ \midrule
\textbf{Physical and MAC}                               & IEEE802.11p (6 Mbit/s)                \\
\textbf{Transmit power}                                 & 100 (mW)                          \\
\textbf{Channel model}                                  & Free space                        \\
\textbf{Beacon size}                                    & 200B                           \\
\textbf{CACC controller}                                & PATH                             \\
\textbf{Total vehicles}                                 & $120 (low density), 480 (high density)$                        \\
\textbf{Vehicles per platoon}                           & $15$                            \\
\textbf{Initial platoon speed}                          & $27.78\, \text{m/s}, (100 km/h)$ \\
$\textbf{JB config:} \boldsymbol{BI_{\textbf{min}}, BI_{\textbf{max}}, p, \Delta u_{\text{max}}}$  & $0.1, 0.4 (s), 1, 2\, (\text{m/s}^2)$ \\
$\boldsymbol{c}$                                 & $0.5\, (\text{m/s}^2)$             \\
$\boldsymbol{k}$                                 & $-5\, (\text{m/s}^2)$             \\
$\boldsymbol{\textbf{$\tau$}}$                 & $20 (s)$                             \\ \bottomrule
\end{tabular}}
\end{table}

\section{Evaluation}
To evaluate JBE, we used PLEXE \cite{b4}, a platooning framework based on two simulators: OMNeT++ \cite{b14}, which simulates the wireless networking component, and SUMO \cite{b15}, which simulates vehicle traffic. In the overall configuration for all experiments, we use an inter-vehicle separation of 5 meters. The distance between platoons is determined by the current leader speed \( \dot{x}_{0} \) and the ACC time headway \( T_h \), which we set to 1.2s (Equation~\eqref{eq:time_headway}). 

\begin{equation}
\label{eq:time_headway}
d_{\text{inter-platoon}} = T_h \cdot \dot{x}_{\text{0}}
\end{equation}

\subsection{Safety}
Safety is the highest priority in platooning. While it is closely linked to string stability, our focus here is to assess safety in challenging scenarios, particularly under high vehicle density. Therefore, we use the global minimum inter-vehicle distance per simulation run as our primary metric which represents the shortest distance between any two platooning vehicles throughout the simulation. This metric effectively captures a scheme’s ability to deliver timely updates, ensuring that vehicle spacing remains as close as possible to the desired distance. The closer the inter-vehicle distance is to the predefined value (5 meters), the safer the platoon is considered. To evaluate this, we conduct the following experiments:

\subsubsection{Follower Slowing Down}
This experiment considers a scenario where the followers rather than the leaders adjust their dynamics. Specifically, at 5s of simulation time, the first followers on one lane change their speed from $100km/h$ (the platoon speed) to $80km/h$. At 20s of simulation time, the followers aim to return to the original speed of the platoon.

\begin{figure}[t]
    \centering
    \begin{subfigure}[t]{0.24\textwidth}
        \centering
        \includegraphics[width=\textwidth]{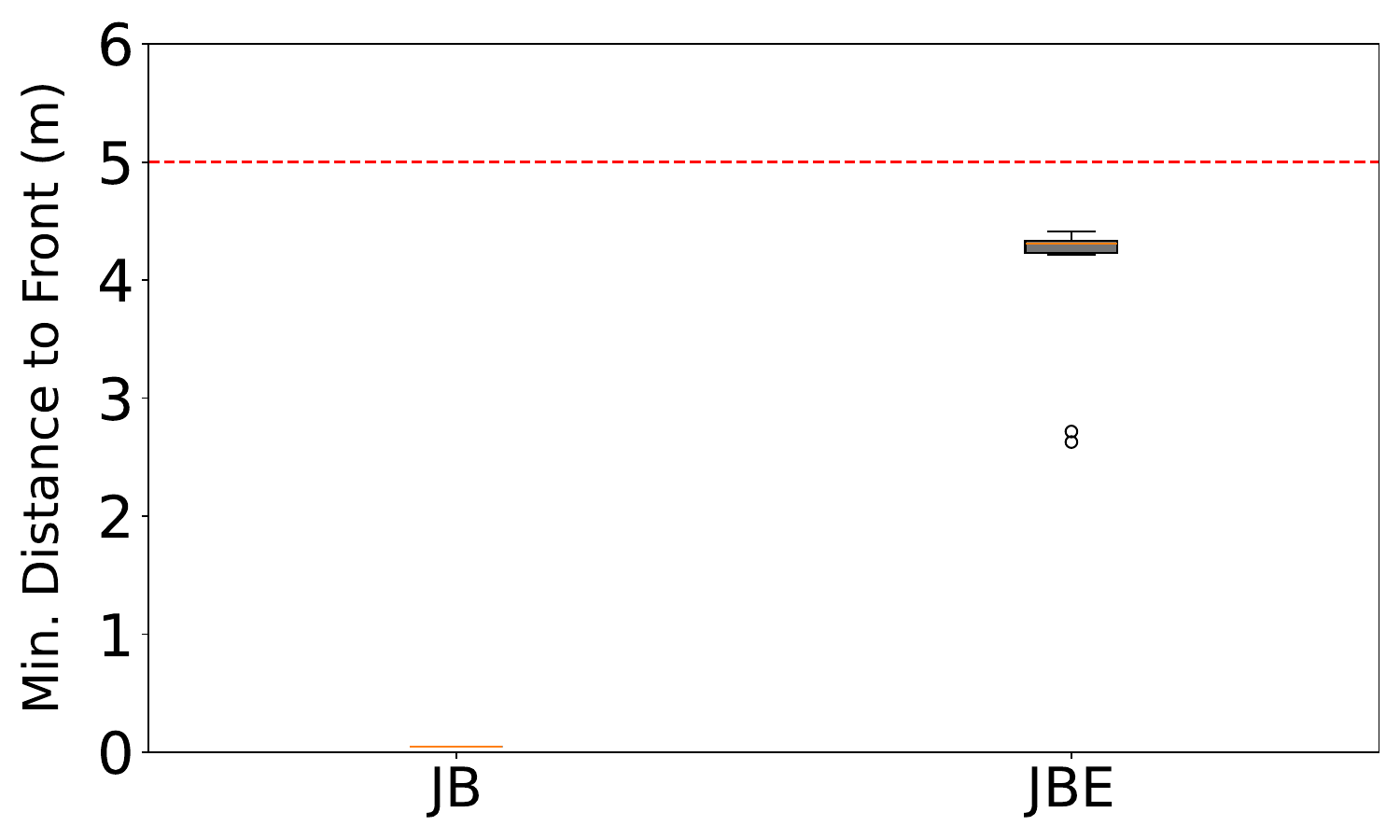}
        \caption{480 vehicles}
        \label{fig:sub1}
    \end{subfigure}
    \hfill
    \begin{subfigure}[t]{0.24\textwidth}
        \centering
        \includegraphics[width=\textwidth]{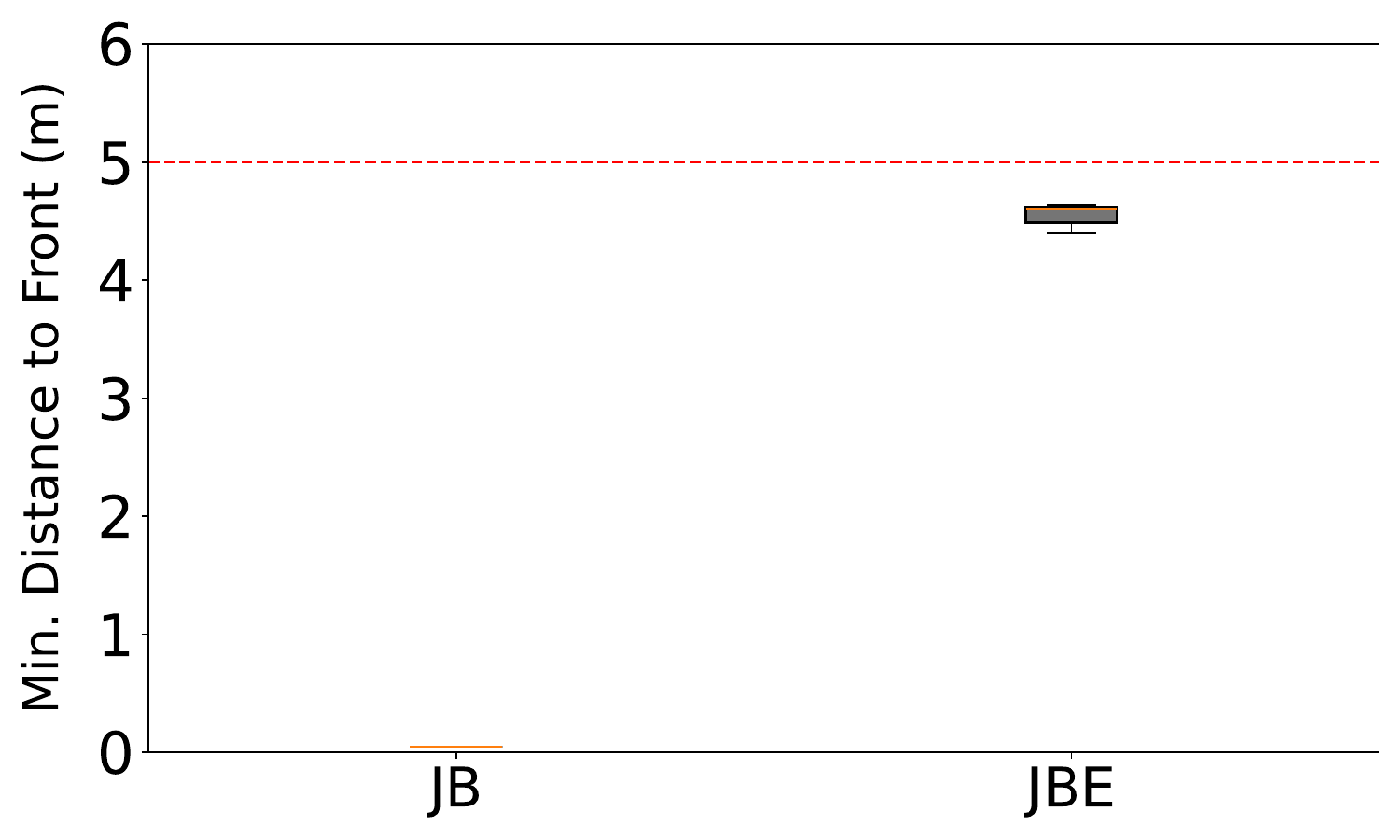}
        \caption{120 vehicles}
        \label{fig:sub2}
    \end{subfigure}
    \caption{Global minimum inter-vehicle distance per simulation run for high and low vehicle densities, 480 and 120 vehicles, respectively. The horizontal dotted red line denotes the target inter-vehicle distance of 5 meters. [Follower Slowing Down].}
    \label{fig:slowing_down_distance}
\end{figure}

As shown in Fig. \ref{fig:slowing_down_distance}, JB consistently led to vehicle crashes in both high and low vehicle densities since it only accounts for deceleration initiated by platoon leaders. In contrast, JBE effectively managed follower dynamics and was able to maintain a safe inter-vehicle distance across both densities. In the low density scenario, inter-vehicle distances consistently remained above 4 meters while in the high-density scenario, there were instances where the distance was about 3 meters. In both cases, no vehicle crashes occurred.

\begin{figure}[t]
    \centering
    \begin{subfigure}[t]{0.24\textwidth}
        \centering
        \includegraphics[width=\textwidth]{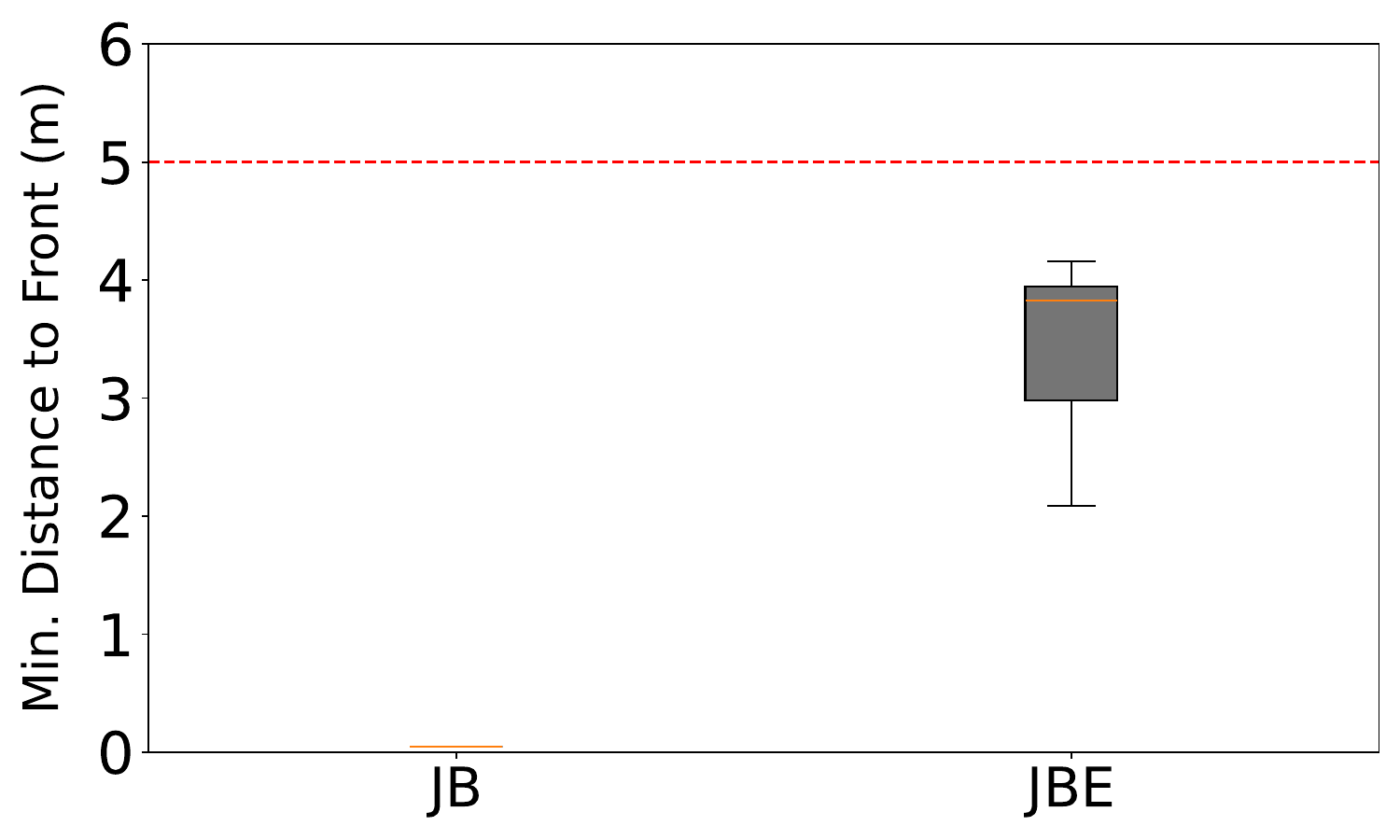}
        \caption{480 vehicles}
        \label{fig:sub1}
    \end{subfigure}
    \hfill
    \begin{subfigure}[t]{0.24\textwidth} 
        \centering
        \includegraphics[width=\textwidth]{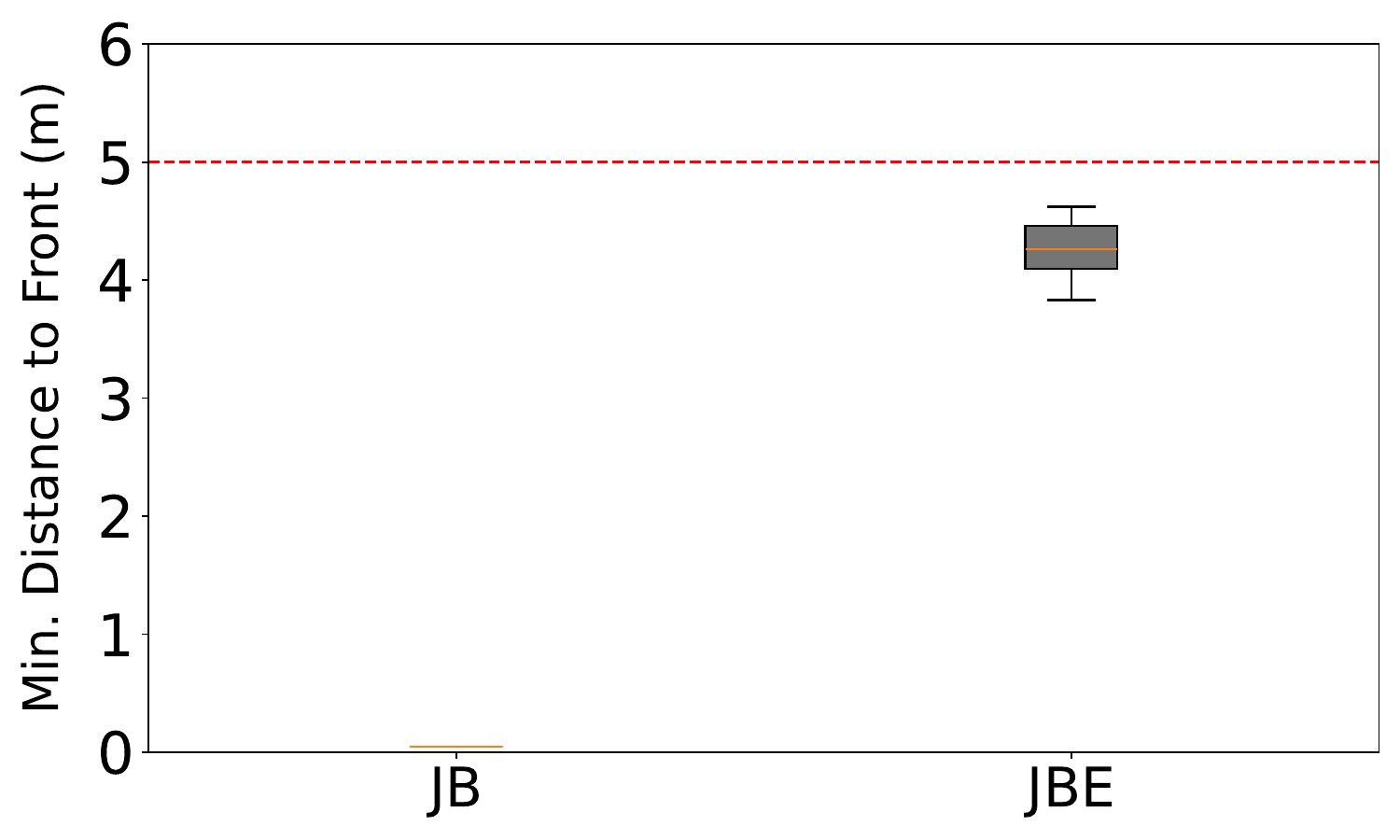}
        \caption{120 vehicles}
        \label{fig:sub2}
    \end{subfigure}
    \caption{Global minimum inter-vehicle distance per simulation run for high and low vehicle densities, 480 and 120 vehicles, respectively. [Follower Stopping].}
    \label{fig:stopping_distance}
\end{figure}

\subsubsection{Follower Stopping}
This experiment also considers a situation in which followers change their dynamics. At 5s of simulation time, the first followers on one lane perform an emergency braking to a complete stop at a constant deceleration rate of $-6 \text{m/s}^2$. Then, at 20s of simulation time, the followers aim to return to the original speed of the platoon ($100km/h$).

Similarly to the previous experiment, as seen in Fig. \ref{fig:stopping_distance}, in both the high and low vehicle densities, JB results in vehicle crashes all the time. JBE, however, was able to handle follower deceleration while maintaining a safe inter-vehicle distance in both scenarios. In the low vehicle density scenario, the distances were always above 4 meters, almost similar to those in the previous experiment. Similarly to the previous experiment, with both vehicle densities there were no vehicle crashes. However, in the high density scenario, there were instances where distance was about 2 meters, yet the median was about 4 meters. From this experiment and the previous one, it is evident that JBE is a safer beaconing scheme than JB.

\subsection{Overhead}
Since the main objective of adaptive beaconing schemes like JB is to reduce the communication channel load, we evaluate the additional overhead introduced by JBE compared to JB to ensure it remains effective in fulfilling its purpose. Therefore, we use the Channel Busy Ratio (CBR) to compare the two schemes in this context. CBR represents the total duration during which the physical layer senses the channel as busy. A low CBR is crucial for minimising delay and packet loss, as the channel is shared among vehicles within the communication range. It also enables other applications, beyond platooning, to use the channel. CBR is calculated using Equation ~\eqref{eq:CBR}, where \( M \) is the total number of fractions of time the channel was sensed busy within the period \( T \) (1s) and \( t_{\text{busy}, i} \) represents the duration for which the channel remained busy during the \( i \)th fraction of time.

\begin{equation}
CBR = \frac{\sum_{i=1}^{M} t_{\text{busy}, i}}{T}
\label{eq:CBR}
\end{equation}

\begin{figure}[t]
    \centering
    \begin{subfigure}[t]{0.24\textwidth}
        \centering
        \includegraphics[width=\textwidth]{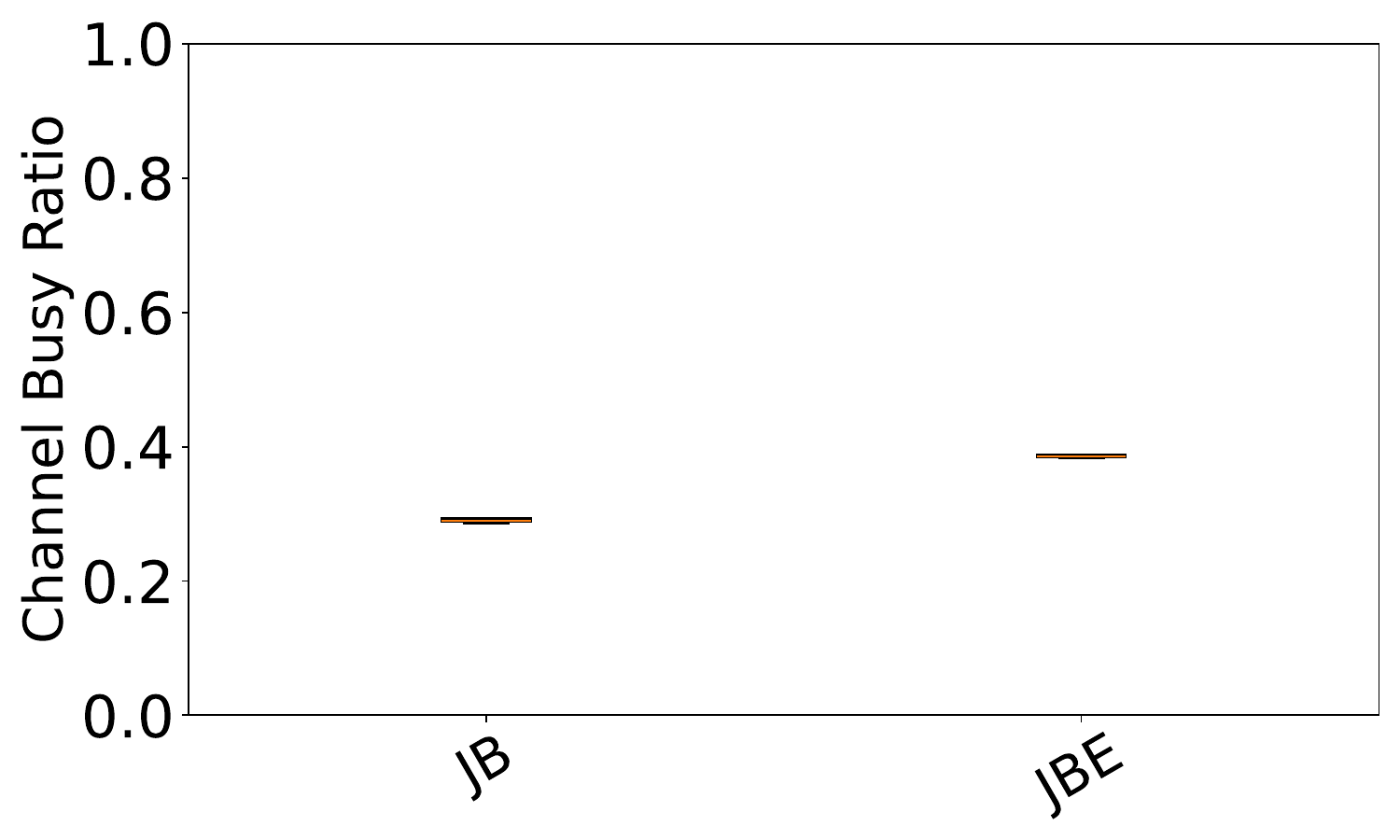}
        \caption{480 vehicles}
        \label{fig:sub1}
    \end{subfigure}
    \hfill
    \begin{subfigure}[t]{0.24\textwidth}
        \centering
        \includegraphics[width=\textwidth]{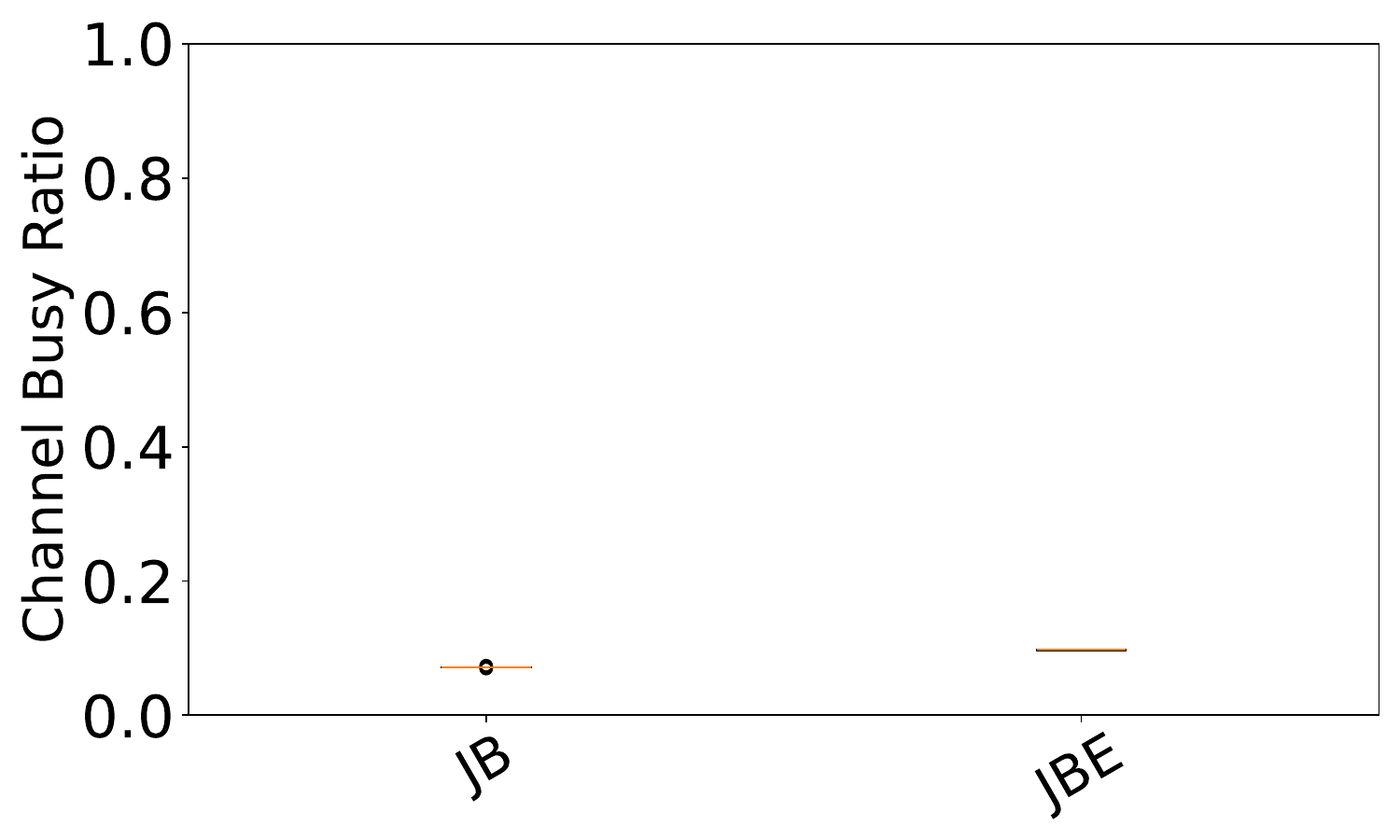}
        \caption{120 vehicles}
        \label{fig:sub2}
    \end{subfigure}
    \caption{Average Channel Busy Ratio (CBR). [Follower Slowing Down].}
    \label{fig:slowing_down_CBR}
\end{figure}

\noindent For the \textit{Follower Slowing Down} experiment (Fig. \ref{fig:slowing_down_CBR}), it is clear that JBE has higher CBR than JB although not significant. In the high vehicle density scenario, JBE has a CBR of about 40\% as compared to that of JB being about 30\%. This is partly because JB always resulted in a vehicle crash, and up to the crash the beaconing interval would always be near maximum.

\begin{figure}[t]
    \centering
    \begin{subfigure}[t]{0.24\textwidth}
        \centering
        \includegraphics[width=\textwidth]{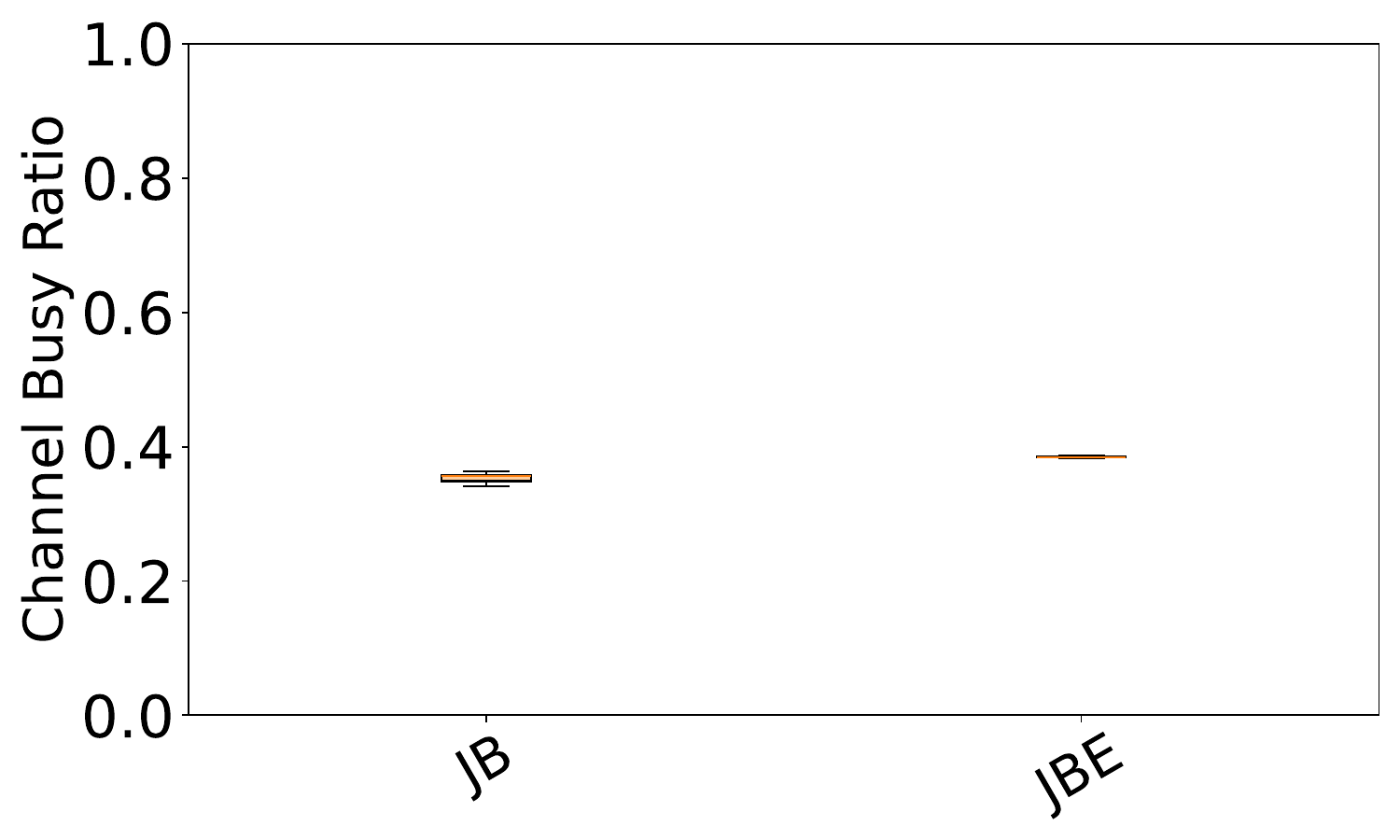}
        \caption{480 vehicles}
        \label{fig:sub1}
    \end{subfigure}
    \hfill
    \begin{subfigure}[t]{0.24\textwidth}
        \centering
        \includegraphics[width=\textwidth]{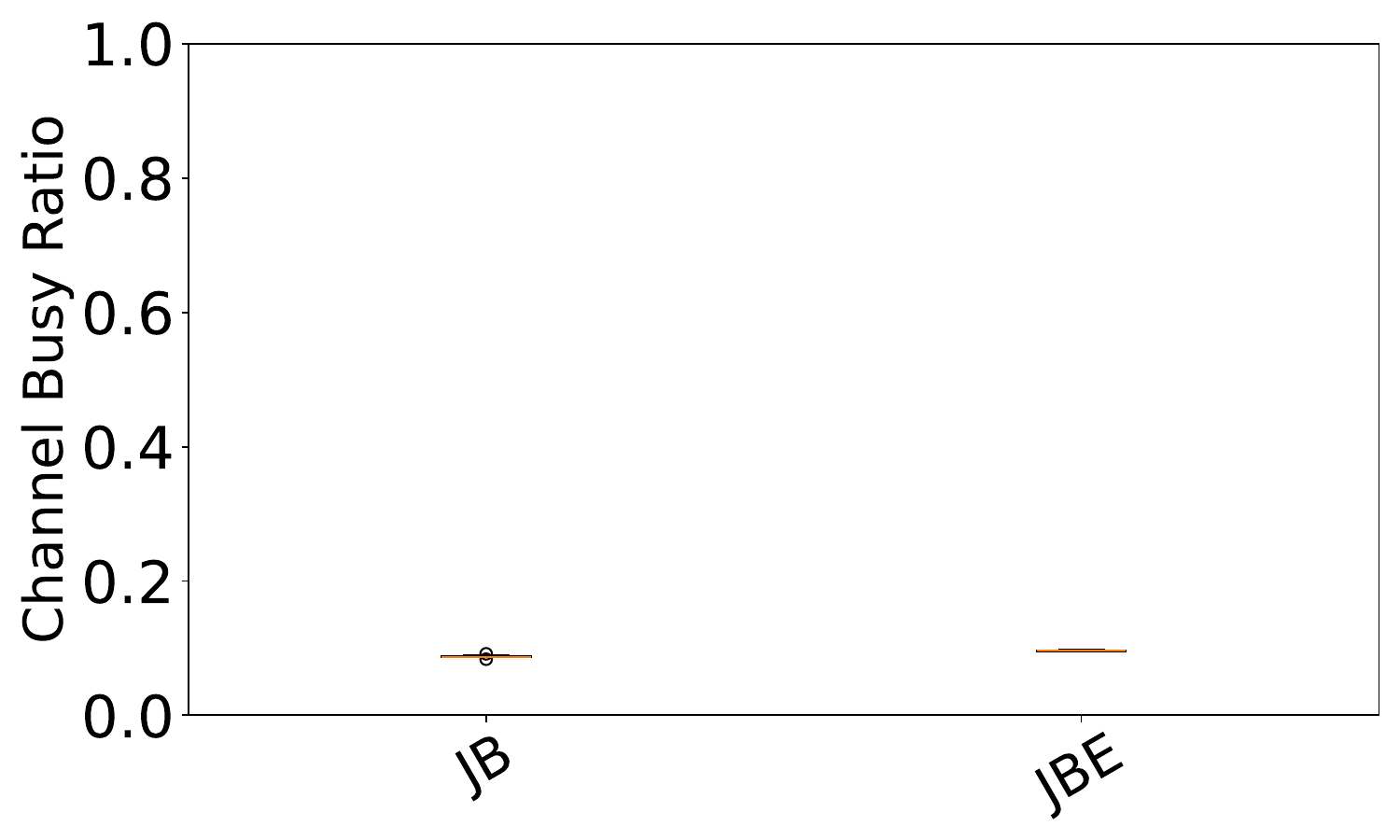}
        \caption{120 vehicles}
        \label{fig:sub2}
    \end{subfigure}
    \caption{Average Channel Busy Ratio (CBR). [Follower Stopping].}
    \label{fig:stopping_CBR}
\end{figure}

Similarly, for the \textit{Follower Stopping} experiment (Fig. \ref{fig:stopping_CBR}), the overhead caused by JBE in terms of CBR was nearly identical to that in the \textit{Follower Slowing Down} experiment which confirms that JBE introduces some overhead to achieve safety. However, we believe that this overhead is acceptable given the safety benefits gained.

\subsection{Impact on string stability}
A key concept in vehicle platooning is string stability, a property that ensures errors in speed or acceleration do not amplify as they propagate through the platoon \cite{b1}. Under normal conditions, PATH CACC (the platooning controller used in this study) is string-stable \cite{b2}. Here, we confirm that JBE does not impact the string stability of this controller. Without loss of generality, we evaluate the impact on stability using the \textit{Follower Stopping} experiment, but with only one platoon of 8 vehicles (a normal condition). In this experiment, two disturbances (changes in speed) occur:
\begin{itemize}
\item The first disturbance (deceleration) is initiated by the first follower at 5s of simulation time.
\item The second disturbance (acceleration) is initiated by the platoon leader at 20s of simulation time.
\end{itemize}

\begin{figure}[t]
    \centering
    \begin{subfigure}[t]{0.24\textwidth}
        \centering
        \includegraphics[width=\textwidth]{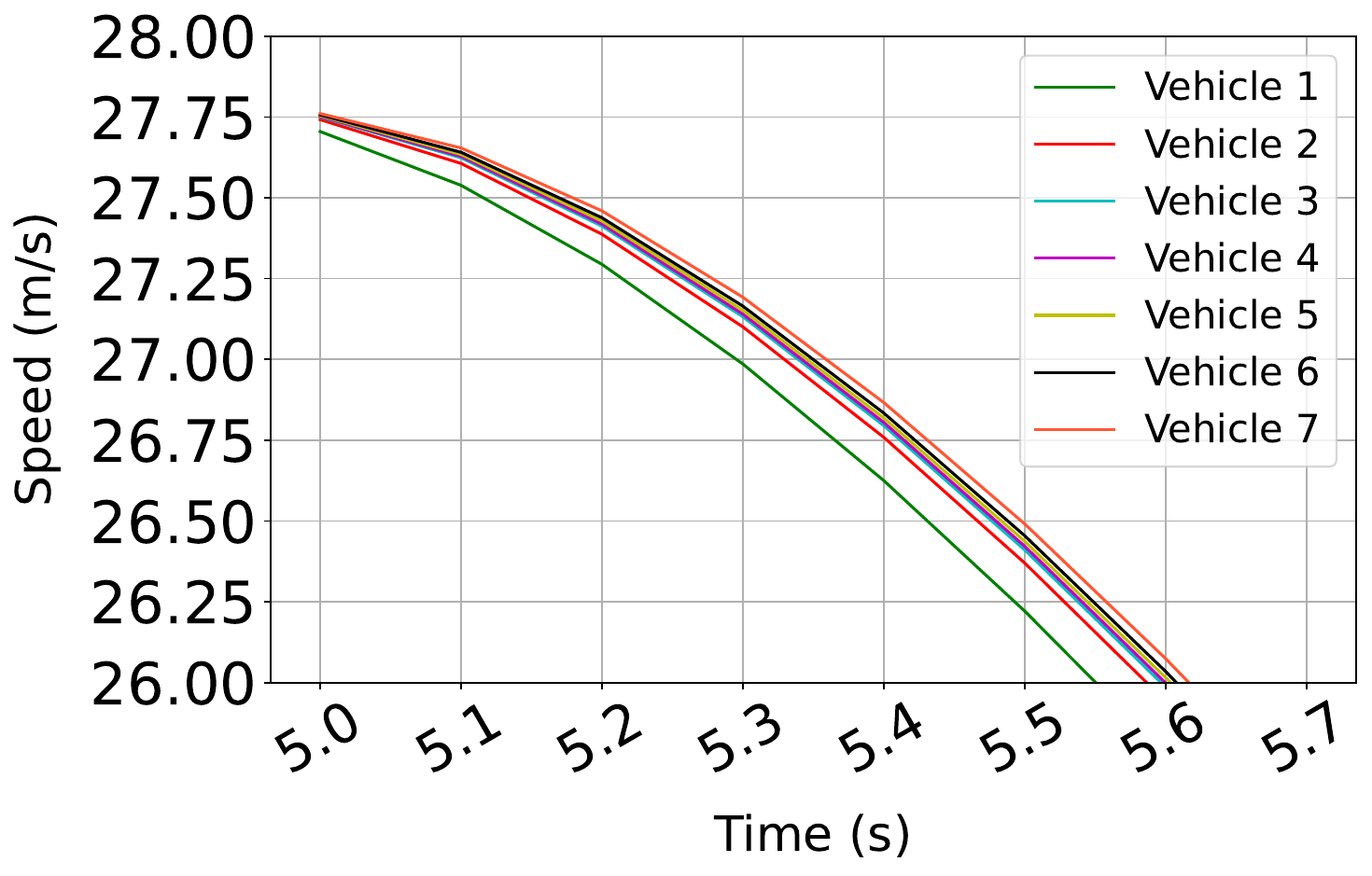}
        \caption{Deceleration}
        \label{string_stability:sub1}
    \end{subfigure}
    \hfill
    \begin{subfigure}[t]{0.24\textwidth}
        \centering
        \includegraphics[width=\textwidth]{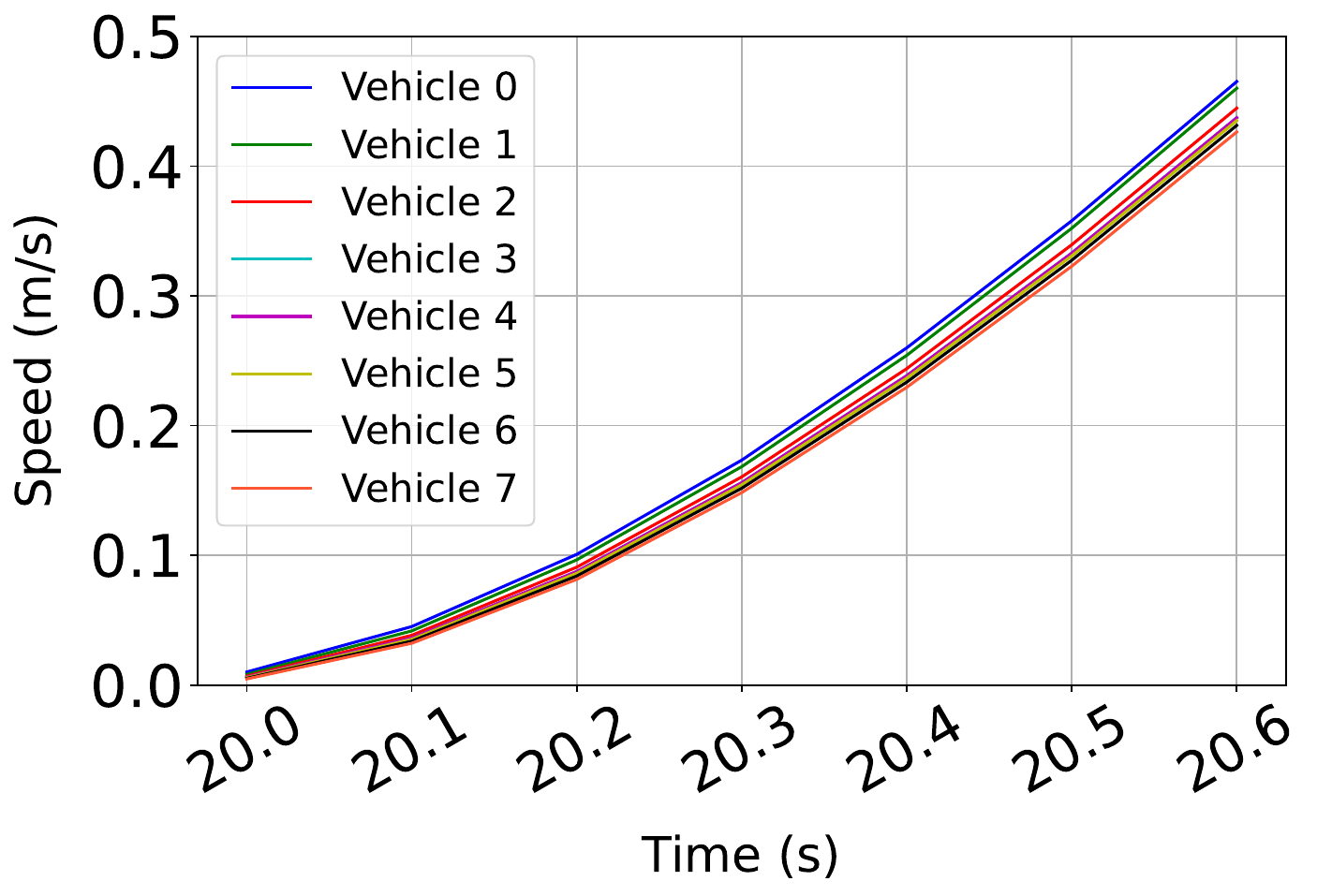}
        \caption{Acceleration}
        \label{string_stability:sub2}
    \end{subfigure}
    \caption{Speeds of platoon members over time during the deceleration and acceleration disturbances which occur at 5s and 20s of simulation time, respectively.}
    \label{fig:string_stability}
\end{figure}

\noindent In Fig. \ref{string_stability:sub1}, the first follower (Vehicle 1) initiates the speed change as indicated by the green line. The following vehicles (red, cyan, etc.) decelerate at a slower rate and this indicates that the disturbance is gradually diminishing. In string stability analysis, the focus is on the vehicle that introduces the disturbance and the ones behind it; therefore, Vehicle 0 is not included in this figure. Considering Fig. \ref{string_stability:sub2}, the platoon leader (Vehicle 0) initiates the speed change (blue line). The following vehicles (green, red, etc.) accelerate at a slower rate and this shows that the disturbance is not being amplified toward the tail of the platoon. Thus, in both cases, the disturbances were not amplified which shows that JBE does not negatively impact the string stability of PATH CACC in normal conditions.

\section{Conclusion}
In this study, we have enhanced the JB beaconing scheme in terms of follower dynamics to improve safety. The results demonstrate that the new version, JBE, significantly enhances safety by addressing follower slowing and braking without introducing substantial CBR overhead. Moreover, JBE does not negatively affect the string stability of the PATH CACC controller in normal conditions.


\begin{thebibliography}{00}

\bibitem{b0}
A. Balador, A. Bazzi, U. Hernandez-Jayo, I. de la Iglesia, and H. Ahmadvand, “A survey on vehicular communication for cooperative truck platooning application,” Vehicular Communications, vol. 35, p. 100460, 2022.

\bibitem{b1}
A. Bose and P. Ioannou, “Analysis of Traffic Flow with Mixed Manual and Intelligent Cruise Control Vehicles: Theory and Experiments,” University of California, Berkeley, California PATH Research Report UCB-ITS-PRR-2001-13, April 2001.

\bibitem{b2}
R. Rajamani, Han-Shue Tan, Boon Kait Law and Wei-Bin Zhang, "Demonstration of integrated longitudinal and lateral control for the operation of automated vehicles in platoons," in IEEE Transactions on Control Systems Technology, vol. 8, no. 4, pp. 695-708, July 2000.

\bibitem{b3}
M. Segata, R. Lo Cigno, H. -M. M. Tsai and F. Dressler, "On platooning control using IEEE 802.11p in conjunction with visible light communications," 2016 12th Annual Conference on Wireless On-demand Network Systems and Services (WONS), Cortina d'Ampezzo, 2016, pp. 1-4.

\bibitem{b4}
M. Segata, S. Joerer, B. Bloessl, C. Sommer, F. Dressler and R. L. Cigno, "Plexe: A platooning extension for Veins," 2014 IEEE Vehicular Networking Conference (VNC), Paderborn, Germany, 2014, pp. 53-60.

\bibitem{b5}
M. van Eenennaam, W. K. Wolterink, G. Karagiannis and G. Heijenk, "Exploring the solution space of beaconing in VANETs," 2009 IEEE Vehicular Networking Conference (VNC), Tokyo, Japan, 2009, pp. 1-8.

\bibitem{b6}
N. Lyamin, A. Vinel, D. Smely and B. Bellalta, "ETSI DCC: Decentralized Congestion Control in C-ITS," in IEEE Communications Magazine, vol. 56, no. 12, pp. 112-118, December 2018.

\bibitem{b7}
G. Bansal, J.B. Kenney, C.E. Rohrs, LIMERIC: a linear adaptive message rate algorithm for DSRC congestion control, IEEE Trans. Veh. Technol. 62 (9) (2013) 4182–4197.

\bibitem{b8}
B. Cheng, A. Rostami, M. Gruteser, H. Lu, J. B. Kenney and G. Bansal, "Evolution of vehicular congestion control without degrading legacy vehicle performance," 2016 IEEE 17th International Symposium on A World of Wireless, Mobile and Multimedia Networks (WoWMoM), Coimbra, 2016, pp. 1-6.

\bibitem{b9}
C. Sommer, S. Joerer, M. Segata, O. K. Tonguz, R. A. Cigno, and F. Dressler, “How shadowing hurts vehicular communications and how dynamic beaconing can help,” IEEE Trans. Mobile Comput., vol. 14, no. 7, pp. 1411–1421, Jul. 2015.

\bibitem{b10}
W. Li, W. Song, Q. Lu, and C. Yue, "Reliable congestion control mechanism for safety applications in urban VANETs," Ad Hoc Networks, vol. 98, p. 102033, 2020.

\bibitem{b11}
M. Segata, B. Bloessl, S. Joerer, C. Sommer, M. Gerla, R. Lo Cigno and F. Dressler, “Toward communication strategies for platooning: Simulative and experimental evaluation,” IEEE Trans. Veh. Technol., vol. 64, no. 12, pp. 5411–5423, Dec. 2015.

\bibitem{b111}
H. Laghbi, N. Thomas, and M. Forshaw, “Slotted and Token-based Overlays for Vehicular Platooning,” in Proceedings of 40th Annual UK Performance Engineering Workshop (UKPEW 2024), University of Leeds, UK, 2024, pp. 4–10.

\bibitem{b12}
L. Hoang, E. Uhlemann, and M. Jonsson, “An efficient message dissemination technique in platooning applications,” IEEE Commun. Lett., vol. 19, no. 6, pp. 1017–1020, Jun. 2015.

\bibitem{b13}
M. Segata, F. Dressler, and R. A. Cigno, “Jerk beaconing: A dynamic approach to platooning,” in Proc. IEEE Veh. Netw. Conf. (VNC), Kyoto, Japan, 2015, pp. 135–142.

\bibitem{b14}
OMNeT++ Community, "OMNeT++ Discrete Event Simulator," [Online]. Available: https://omnetpp.org/.

\bibitem{b15}
P. A. Lopez, M. Behrisch, L. Bieker-Walz, J. Erdmann, Y.-P. Flötteröd, R. Hilbrich, L. Lücken, J. Rummel, P. Wagner, and E. Wießner, “Microscopic traffic simulation using SUMO,” in Proc. 21st Int. Conf. Intell. Transp. Syst. (ITSC), Maui, HI, USA, 2018, pp. 2575–2582.

\bibitem{b16}
M. Amoozadeh, H. Deng, C.-N. Chuah, H. M. Zhang, and D. Ghosal, "Platoon management with cooperative adaptive cruise control enabled by VANET," Vehicular Communications, vol. 2, no. 2, pp. 110–123, 2015.

\bibitem{b17}
H. Laghbi and N. Thomas, "Performance Evaluation of Beaconing Schemes for Vehicular Platooning," in Proceedings of the European Performance Engineering Workshop (EPEW), vol. 15454, Lecture Notes in Computer Science, Cham, Switzerland: Springer, 2024, pp. 29–44.



\end{thebibliography}
\end{document}